# Dragon Crypto – An Innovative Cryptosystem

Awnon Bhowmik
Department of Mathematics and Computer Science
CUNY York College
94 – 20 Guy R. Brewer Blvd
Jamaica, NY 11451

Unnikrishnan Menon
Department of Electrical and Electronics Engineering
Vellore Institute of Technology, Vellore
Tamil Nadu 632014

## ABSTRACT
In recent years cyber-attacks are continuously developing. This means that hackers can find their way around the traditional cryptosystems. This calls for new and more secure cryptosystems to take their place. This paper outlines a new cryptosystem based on the dragon curve fractal. The security level of this scheme is based on multiple private keys, that are crucial for effective encryption and decryption of data. This paper discusses, how core concepts emerging from fractal geometry can be used as a trapdoor function for this cryptosystem.

## 1. INTRODUCTION
Fractals are just a self-replication of a pattern. There are many fractals that are found in nature. One such fractal curve is known as *Heighway Dragon*, or simply the *Dragon Curve*. Because this is a fractal, hence it requires an IFS (Iterated Function System). In simple words, an initial simple pattern is made and thrown into a recursive function, before running a loop and calling the function repeatedly until the required objective is accomplished. The idea of this cryptosystem came up due to the blog post (Bhowmik & Menon, 2018) about coding up a dragon curve fractal and recently while playing around with twin dragon, tetra dragon and septa dragon curves, which was generated as a fun part of this project and named the rainbow dragon (Bhowmik & Menon, Rainbow Dragon, 2020) since each part was given a color of the VIBGYOR.

## 2. THE DRAGON FRACTAL
A fractal is just a repetition of an initial geometric shape. This is generated by something known as the Iterative Function System (IFS). The dragon curve is one such fractal. There are multiple ways to generate this fractal.

### 2.1 Generating the Dragon Fractal
This dragon curve fractal, or its generation algorithm forms the basis of this cryptosystem, and hence the proposed name *Dragon Crypto*. Here, an arbitrary string of characters is passed through the *Koblitz Encoder* to obtain the starting point for the IFS.

The iterations of the dragon curve can easily be generated by folding a strip of paper $n$ number of times. A strip of paper is taken and folded in half, to the right. This process is repeated, folded in half again to the right. The process is continued as many times as required, which is usually many times. It can be easily noticed that it will be hard to fold the strip after a certain point. Suppose the strip is folded twice. When unfolded and relaxed, it is observed that every bend/corner of the strip has a 90° turn. It is now the second iteration of the dragon curve. If it is folded again and opened, it would be the third iteration and so on. Following is a visual of what is described here…

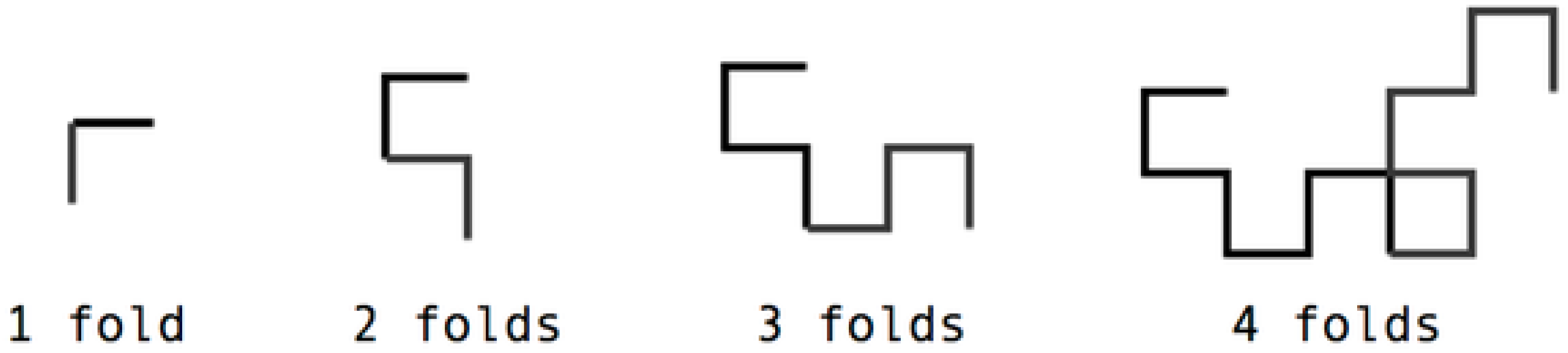

**Figure 1. Paper folding method**

Now all that's left to do is to visualize a turtle walking along these lines and it should be able to predict how to move. Let F = forward, L = left, R = right. The following sequences are associated with the respective iterations…

- 1st iteration – F L F L
- 2nd iteration – F L F L F R F L
- 3rd iteration – F L F L F R F L F L F R F R F L

If the $n^{th}$ iteration is known, then the $(n+1)^{th}$ iteration can be predicted in the following way. Suppose the objective is to derive the third iteration from the second iteration

- The last element of the sequence if ignored for the moment (F L F L F R F). In the remaining elements, the L's and R's are switched. The sequence now becomes (F R F R F L F R)
- Now the elements about the midpoint are flipped.

The sequence turns into (R F L F R F R F)

- The last element which was ignored in the first step is appended to the sequence (R F L F R F R F L)
- Appending the result of step 4 to the 1st iteration will generate the 3rd iteration.

Using the same logic, the next iterations can be predicted. A fractal curve generated with 15 iterations is as follows…

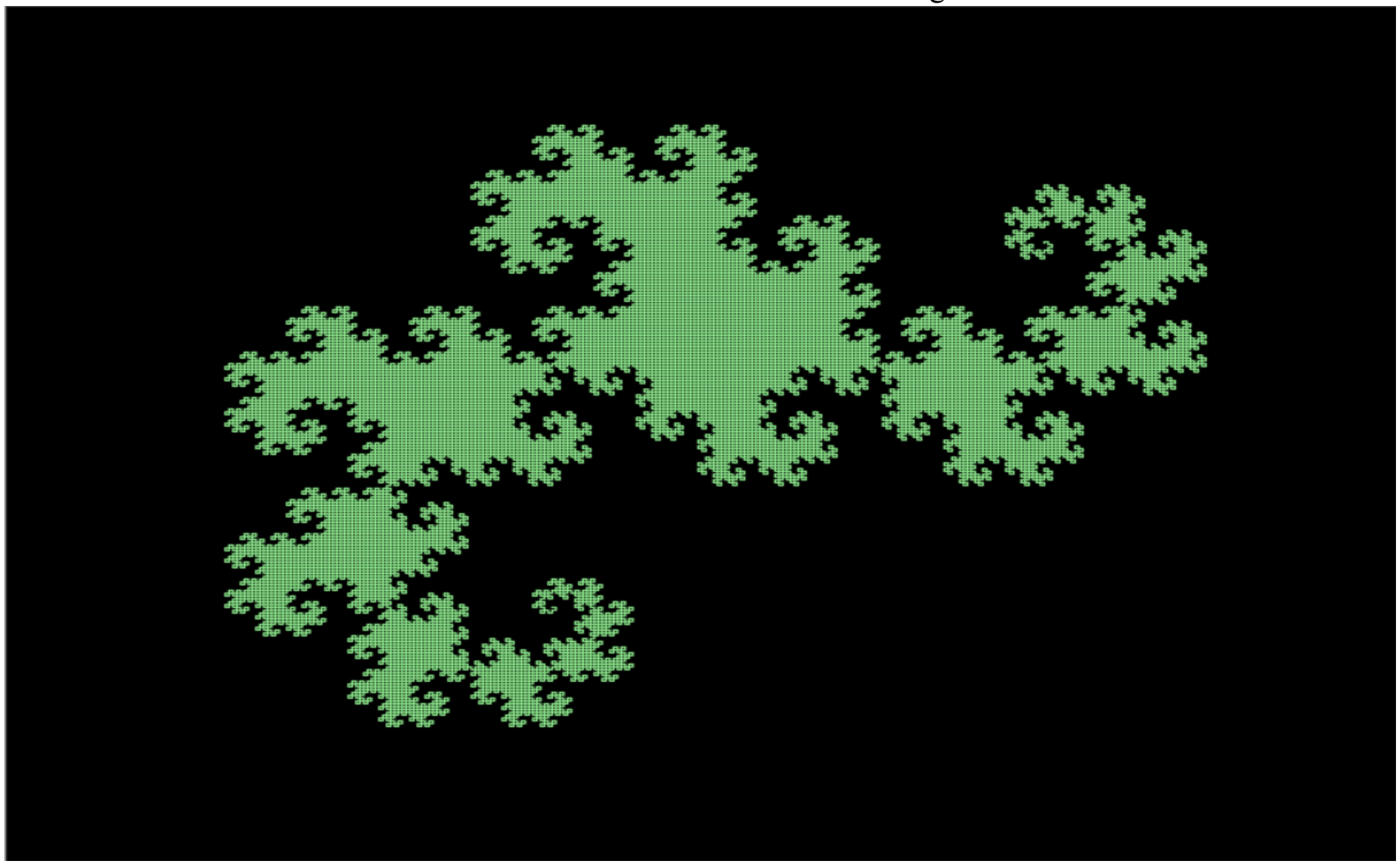

**Figure 2. Dragon Curve Fractal**

## 3. KOBLITZ ENCODING AND DECODING ALGORITHM

The encoding algorithm (Brady, Davis, & Tracy, 2010) is as follows

- Given a message $M$, convert each character $m_k$ into a number $a_k$ using Unicode, where $b = 2^{16}$ and $0 < a_k < 2^{16}$
- Convert the message $M$ into an integer using

$$m = \sum_{k=1}^{n} a_k b^{k-1}$$

In practice an $n \leq 160$ is chosen such that $m$ satisfies

$$m \leq 2^{16 \cdot 160} < p$$

- A number $d$ is fixed such that $d \leq \frac{p}{m}$. In practice the prime $p$ is chosen large enough so that $d = 100$ can be allowed.
- For integers $j = 0,1,2, \ldots d - 1$, the following are performed
- $x$ coordinate of a point on the elliptic curve is computed as

$$x_j = (dm + j) \bmod p \text{ where } m = \left\lfloor \frac{x_j}{d} \right\rfloor$$

- Compute $s_j = \left(x_j^3 + Ax + B\right) \bmod p$
- If $\left(s_j\right)^{\frac{p+1}{2}} \equiv s_j \bmod p$, then $y$ coordinate of a point on the elliptic curve is defined as $y_j = \left(s_j\right)^{\frac{p+1}{4}} \bmod p$. Return the point $\left(x_j, y_j\right)$.

Thus, the message $M$ is encoded as an element of the Abelian group $G = E\left(\mathbb{F}_p\right)$. The following is performed for decryption.

- Considering each point $(x, y)$ and setting

$$m = \left\lfloor \frac{x - 1}{k} \right\rfloor$$

Which is essentially means $a_k = \left\lfloor \frac{m}{b^{k-1}} \right\rfloor \bmod b$. Thus, each character is recovered and concatenated to produce the original message $M$.

## 4. EQUIVALENCE CLASS MODULO 4

Any number divided by 4 can result in remainders {0,1,2,3}. So, any number divided by 4 is one of $4r, 4r + 1, 4r + 2, 4r + 3$. However, if only odd numbers are considered, they are of the form $4r + 1$ or $4r + 3$. Numbers of these forms are useful in figuring out important information regarding the dragon curve and could be an element of penetration attempt by an interceptor. This is described in detail at a later section.

**Theorem 1.** Let $R \subseteq S \times S$ be an equivalence class on a set $S$. Then the set of $\mathcal{R}$-classes constitutes the whole of $S$.

*Proof.*

| | |
|---|---|
| $\forall x \in S: x \in [x]_{\mathcal{R}}$ | Definition of equivalence class |
| $\neg(\exists x \in S: x \notin [x]_{\mathcal{R}})$ | Assertion of universality |
| $\neg(\exists x \in S: x \notin \cup [x]_{\mathcal{R}})$ | Definition of set union |
| $\forall x \in S: x \in \cup S/\mathcal{R}$ | Assertion of universality |
| $S \subseteq \cup S/\mathcal{R}$ | Definition of a subset |

Also:

| | |
|---|---|
| $\forall X \in S/\mathcal{R}: X \subseteq S$ | Definition of equivalence class |
| $\cup S/\mathcal{R} \subseteq S$ | Assertion of universality |

By definition of set equality

$$\cup S/\mathcal{R} \subseteq S$$

And so, the set of all $\mathcal{R}$-classes constitutes the whole of $S$. (Union of Equivalence Classes is Whole Set, n.d.)

# 5. TRAPDOOR FUNCTION

A trapdoor function is a highly useful concept in modern cryptography. These are functions that are easy to compute in one direction but extremely hard to compute in reverse if certain parameters or critical information for reversal is lacked. The main novelty of this cryptosystem is the use of the *Heighway Dragon Fractal* as a trapdoor function. The algorithm starts off with a secret message that needs to be encrypted (called the plainText).

The *Koblitz Encoder* accepts the plainText along with the curve parameters (obtained from the private key). The Koblitz Encoder spits out an encoded point in 2D Cartesian space for each character present in the plainText with the help of curve parameters. These points are now the starting point of the dragon curve fractals. For each character, a corresponding starting point and a dragon fractal are generated.

Now the components of the private key (including size, iterations, angle) are used to generate the fractal for each character from their corresponding starting points.

## 5.1 Private key parameters

The following are the private keys involved in the cryptographic algorithm

1. The size defines the length of each forward step
2. The number of iterations for generating the fractal
3. The starting angle for the fractal
4. Elliptic curve parameters $a$ and $b$ for the function $y^2 = x^3 + ax + b$

Once the fractals are generated for each encoded starting point, the corresponding endpoints are noted and stored.

$$p_{start} = \{(x_i, y_i)\}, \quad i = 1,2,\dots,n$$
$$p_{end} = \{(x_j, y_j)\}, \quad j = 1,2,\dots,n$$

# 6. PROPOSED ALGORITHM

1. Input an arbitrary string, and split into list of characters, spaces and special characters

2. **Encryption**

a. Each character's ASCII representation is passed through the *Koblitz Encoder* (using Koblitz algorithm in section ). Every character is then represented by a point $(x_i, y_i)$ on the Cartesian coordinate. The set of these points are initial starting points.

b. For a fixed length $l$, an angle $\theta$ and a fixed number of iterations $n$ for all characters (or points) in the set, a dragon fractal is generated. The end point $(x_j, y_j)$ is stored into another list.

c. A special padding is applied on the list of end points as follows

$$"Xx_1Xx_2 \cdots Xx_nXYy_1Yy_2 \cdots y_nY"$$

This turns it into a string. This string is the encrypted cipher text.

3. **Decryption**

a. Padding is removed and points are regenerated. The string is parsed through and split into two sequences about the "$XY$" mark. So now there are two strings

$$Xx_1Xx_2 \cdots Xx_nX \quad \text{and} \quad Yy_1Yy_2 \cdots y_nY$$

Next, the ordered set $(x_j, y_j)$ of endpoints are recovered by parsing through the two strings.

b. The list of end point coordinates is taken and based on the parameters present in the private key (size, iterations, angle), a dragon curve fractal is generated in reverse by initializing the turtle at an endpoint facing the correct direction based on the angle. If the private key parameters are all correct, this reverse tracing means the fractal will end at the initial starting point. This process is repeated for all characters. The points obtained are stored in a list.

c. The list of points obtained from previous step is passed through the *Koblitz Decoder*. This returns the list of ASCII values of each character in the original string. This is then converted to list of characters and joined to retrieve the original string.

# 7. PROBABLE IMPLICABLE TACTICS BY INTERCEPTOR

## 7.1 Determining the turn in the fractal

The number of iterations $n$ in the algorithm described above is a private key. If somehow an interceptor can obtain this value, they can very easily determine the left or right turn in any of those $n$ iterations. This gives them an idea about the trajectory taken by the fractal for a certain character in the original message during the encryption procedure. The turn determination is done by doing the following

- $n = 2^m k$, where $k$ is obviously odd
- $k \bmod 4 = \begin{cases} 1, & k = L \\ 3, & k = R \end{cases}$

which is known as the **left right rule** (Geometry). But they would need to generate each turn, feed it to a list and make the turtle follow that sequence. For example, a sequence of $\{0,1,2,3\} = \{F, L, F, R\}$. Notice that 0 and 2 corresponds to forward movement in the already set direction only. But for a given large number of iterations, it is hard to determine where the curve originated from since the interceptor lacks the starting encoding values $(x_i, y_i)$ of the characters to be encrypted.

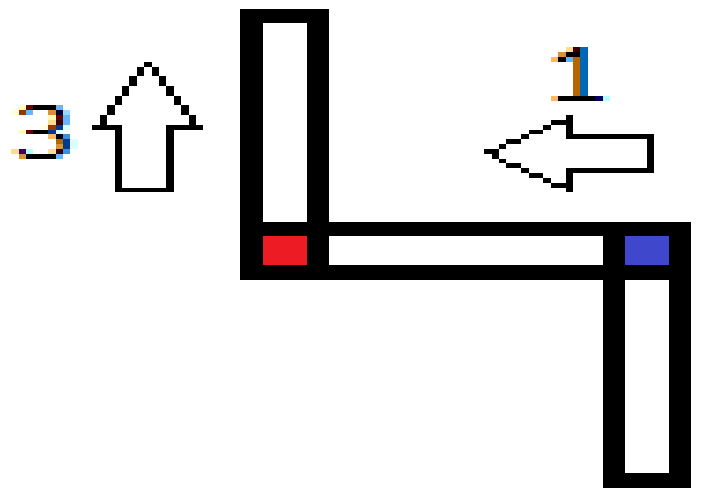

**Figure 3. Determining the turn in fractal**

## 7.2 Variation in the fractal parameters

If the list of starting and ending points of the characters, is available to the interceptor, he cannot generate the exact fractal connecting the two points since the length, angle and the number of iterations is unknown. It can also be that someone can tweak the existing program such that the length or the angle switches to a different value after a certain number of iterations, which could be regarded as a form of "period", making it quite impossible for the interceptor.

## 7.3 Euclidean distance as a public key

Consider the scenario consisting of Alice and Bob with Eve being the eavesdropper, or interceptor according to this section so far. Eve knows the angle and the Euclidean distance between the starting points (plain text) and the ending points (cipher text). Eve still won't be able to initiate decryption procedure since she lacks knowledge of number of iterations and step size (length), since the fractal can be generated in

1. Keeping the length small, and the iterations relatively large
2. Keeping the length large, and the iterations relatively small

If Eve does not have access to the right combination of length and the number of iterations, decryption would be impossible

```
---------------------------------Dragon_Crypto-----------------------------------

Enter Message: HELLO

----------------------------------Private Key------------------------------------
Size of Dragon: 5
No. of Iterations: 3
Angle: 0
Curve Parameters
Enter A: -1
Enter B: 7

----------------------------------ASCII Values-----------------------------------

[72, 69, 76, 76, 79]

---------------------------------Koblitz Points----------------------------------

[(1441, 173), (1382, 247), (1522, 59), (1522, 59), (1581, 204)]

---------------------------------Padded Encryption-------------------------------

Encrypted Message:  X1431X1372X1532X1532X1571XY183Y237Y49Y69Y214Y

---------------------------------Split String------------------------------------

X1431X1372X1532X1532X1571X      and      Y183Y237Y49Y69Y214Y

---------------------------Padding Removed and Regenerate-------------------------

[(1431, 183), (1372, 237), (1532, 49), (1532, 69), (1571, 214)]

------------------------------------Decryption------------------------------------

Decrypted message:  HELLO
```

**Figure 4. Sample Test Run**

# 8. RECURSIVE METHOD OF FRACTAL GENERATION

Even though section 2 explains a way to generate a fractal, there are recursive mathematical equations or IFS (Bourke, Macintosh IFS manual, 1990), that can map each coordinate of a fractal given an initial set of parameters such as

$$x_{n+1} = ax_n + by_n + c$$

$$y_{n+1} = dy_n + ex_n + f$$

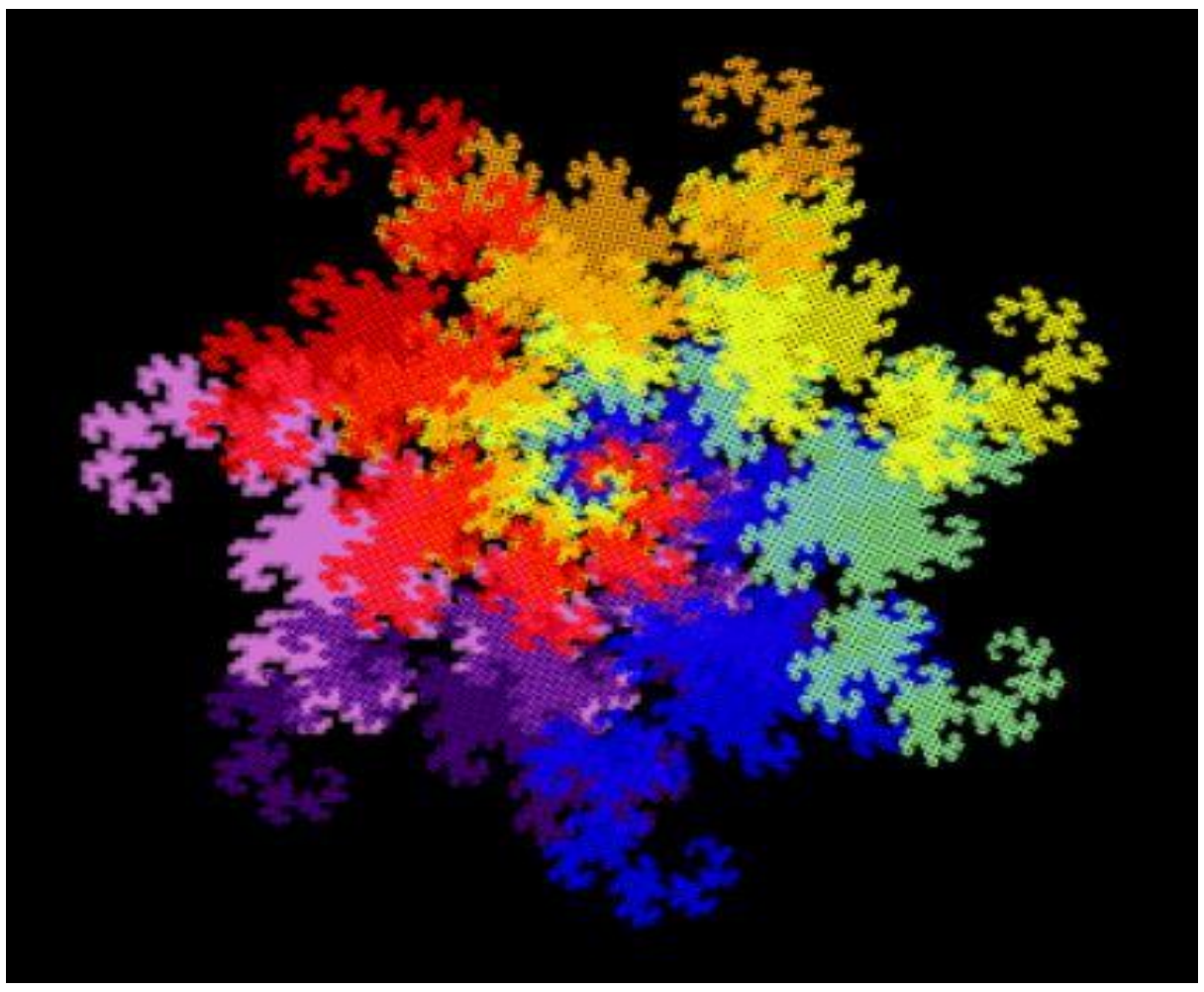

**Figure 5. Rainbow Dragon with IFS**

# 9. EXPERIMENTAL ANALYSIS

Following is a table depicting the time taken to complete encrypt-decrypt cycle for different plaintext sizes. Only a few values are shown in the table, but the graph shows all the points.

**Table 1. Cycle time for varying plaintext size**

| Plaintext Length | Time for encrypt-decrypt cycle |
|---|---|
| 26 | 6.41 |
| 1066 | 279.69 |
| 2106 | 626.09 |
| 3146 | 1106.89 |
| 4186 | 1577.13 |
| 5226 | 2019.75 |
| 6266 | 2794.44 |

The execution time of the application can be shortened if the rendering of the fractal generation is turned off on the viewport.

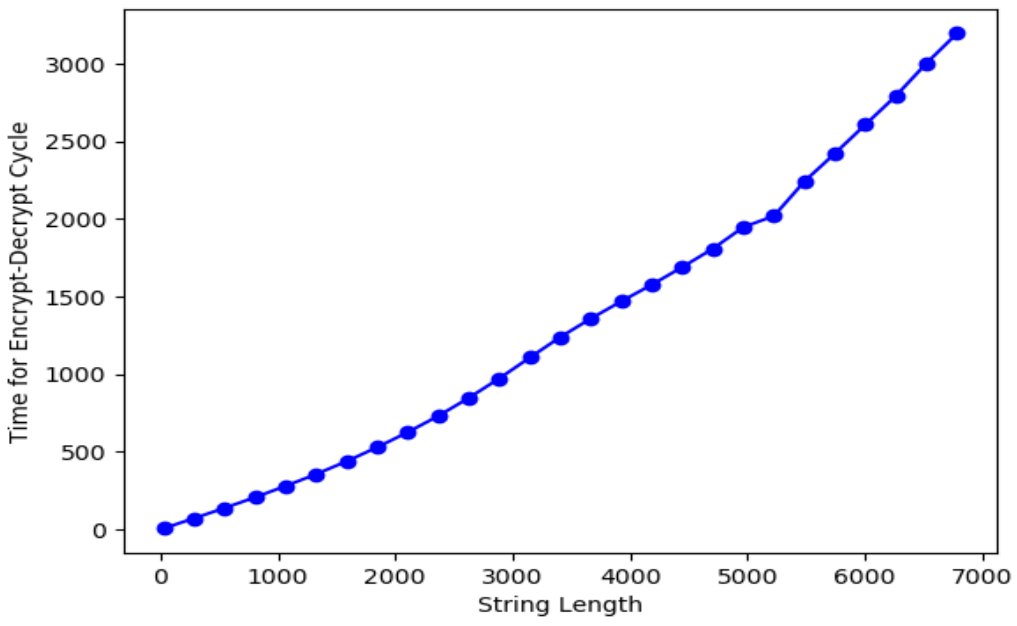

**Figure 6. String Length vs. Cycle Time**

Experiments were performed by varying the elliptic curve parameters $A, B \in [-10,10]$. The fractal parameters were kept constant and the plaintext string consisted of randomly shuffled characters that are found on a standard English keyboard. It was observed that for all combinations of the curve parameters, the decrypted message perfectly matched with the original plaintext without any errors.

# 10. CONCLUSION

In this paper, a new encryption algorithm was proposed based on the well-known dragon curve fractal. Cryptography is a 40 year old topic where a lot has been discovered but a lot more yet remains unknown. The essence of any traditional cryptosystem relies on three hard mathematical problems: the integer factorization problem, **the discrete logarithm problem** (Silverman, 2006) or the elliptic curve discrete logarithm problem. Shor's algorithm can be used to easily compromise the security of these conventional cryptosystems. So, for a secure future, it is critical to come up with innovative trapdoor functions that can be incorporated in the heart of the encryption scheme.

The proposed algorithm is working based on the appropriate parameters. It has been noticed that the run time complexity can be drastically reduced by using smaller number of iterations while increasing the length to compensate for the precision of the endpoint as well as the Euclidean distance from the start point. So far, no edge cases have been found, but the presence of one or more is suspected. This is dependent on the calculation precision of the machine. The work described in this paper is available here (Bhowmik & Menon, Dragon-Crypto, 2020)